\documentclass{PoS}
\usepackage{amsmath}
\usepackage{graphicx}

\title{Measurement of thermodynamics using gradient flow}

\ShortTitle{Measurement of thermodynamics using gradient flow}

\author{\speaker{Masakiyo Kitazawa}\\
        Department of Physics, Osaka University, 
        Toyonaka, Osaka 560-0043, Japan\\
        E-mail: \email{kitazawa@phys.sci.osaka-u.ac.jp}}

\author{Masayuki Asakawa\\
        Department of Physics, Osaka University, 
        Toyonaka, Osaka 560-0043, Japan\\
        E-mail: \email{yuki@phys.sci.osaka-u.ac.jp}}

\author{Tetsuo Hatsuda\\
        Theoretical Research Division, Nishina Center, RIKEN, Wako
        351-0198, Japan,\\
        Kavli IPMU (WPI), The University of Tokyo, Chiba 606-8502, Japan \\
        E-mail: \email{thatsuda@riken.jp}}

\author{Takumi Iritani\\
        Yukawa Institute for Theoretical Physics, Kyoto
        606-8512, Japan\\
        E-mail: \email{iritani@yukawa.kyoto-u.ac.jp}}

\author{Etsuko Itou\\
        High Energy Accelerator Research Organisation (KEK), Tsukuba
        305-0801, Japan\\
        E-mail: \email{eitou@post.kek.jp}}

\author{Hiroshi Suzuki\\
        Department of Physics, Kyushu University, 6-10-1 Hakozaki,
        Higashi-ku, Fukuoka, 812-8581, Japan\\
        E-mail: \email{hsuzuki@phys.kyushu-u.ac.jp}}

\abstract{
We analyze bulk thermodynamics and correlation functions of the
energy-momentum tensor in pure Yang-Mills gauge theory 
using the energy-momentum tensor defined by the gradient flow 
and small flow time expansion.
Our results on thermodynamic observables are consistent with 
those obtained by the conventional integral method.
The analysis of the correlation function of total energy supports 
the energy conservation. 
It is also addressed that these analyses with gradient flow 
require less statistics compared with the previous methods.
All these results suggest that the energy-momentum tensor can
be successfully defined and observed on the lattice with moderate
numerical costs with the gradient flow.
}

\FullConference{The 32nd International Symposium on Lattice Field Theory,\\
		23-28 June, 2014\\
		Columbia University New York, NY}

\begin{document}

\section{Preliminary}
\label{sec:intro}

In this manuscript, we consider the energy-momentum tensor
\begin{align}
  T_{\mu\nu}. \nonumber
\end{align}
As is well known, the energy-momentum tensor (EMT) is one of the
most fundamental quantities in physics.
It is fundamental because this operator is related to the symmetry
of space-time; the EMT can be defined as the Noether current
of translational symmetry, and it also constitutes generators
of Poincar\'e transformation.
Moreover, elements of the EMT, energy and momentum densities and
stress tensor, are basic quantities in physics. The EMT also 
appears in various fundamental equations such as Einstein
equation and hydrodynamic equations.

However, if one wants to define and analyze the EMT in 
lattice QCD Monte Carlo simulations, there are highly nontrivial
problems.
First, the definition of the EMT itself is nontrivial on the lattice,
because the lattice regularization explicitly
breaks translational symmetry.
For example, a na\"ive definition,
\begin{align}
  T_{\mu\nu} = \frac1{g_0^2} F_{\mu\rho}^a F_{\nu\rho}^a
  - \frac1{4g_0^2} \delta_{\mu\nu} F_{\rho\sigma}^aF_{\rho\sigma}^a,
  \label{eq:Tdef}
\end{align}
with a choice of discretized definitions for the field
strength $F_{\mu\nu}$ on the lattice, does not give the 
correct EMT even for off-diagonal components in the sense that
it does not correspond to the EMT in the continuum limit 
\cite{Caracciolo:1989pt}.
Although the EMT on the lattice can be constructed 
by an adequate linear combination of operators, 
the renormalization factor of each operators has to be 
determined non-perturbatively, which requires additional 
numerical simulations for each lattice parameter 
\cite{Caracciolo:1989pt,Meyer:2011gj}.
Second, even if one employs a definition of the EMT
on the lattice, its measurement is noisy.
For example, the calculation of the correlation function
of the EMT requires extremely large statistics 
\cite{Nakamura:2004sy,Meyer:2011gj}.
The analysis becomes more noisy as the lattice spacing
becomes finer.
Because of these reasons, there are not many studies which 
have analyzed the EMT directly on the lattice.

On the other hand, if one were able to deal with the EMT 
on the lattice in a controllable manner, various interesting 
analyses would become possible, because the EMT is one of the
most fundamental quantities in physics.
For example, if we have the correctly normalized EMT operator 
on the lattice one can calculate thermodynamic quantities, 
energy density $\varepsilon$ and pressure $P$, at nonzero 
temperature by directly taking the thermal expectation values 
of the diagonal components of the EMT as 
\begin{align}
  \varepsilon = -\langle T_{00} \rangle , ~ P = \langle T_{ii} \rangle.
  \label{eq:ep}
\end{align}
The measurement of bulk thermodynamics on the lattice itself 
can be done without the EMT; the method based on the 
thermodynamic relations called the integral method 
\cite{Boyd:1996bx} is well established.
The measurement with simple relations Eq.~(\ref{eq:ep}), however, 
will provide us with more intuitive way to investigate these quantities.
Next, the fluctuations and correlation functions of
the EMT are also interesting observables.
The fluctuation of the total energy of the system 
$\bar{T}_{00}$, for example,
is related to the specific heat per unit volume as \cite{LL}
\begin{align}
  \frac{c_V}{T^3} = \frac{\langle \delta \bar{T}_{00}^2 \rangle}{VT^5}.
  \label{eq:c_V}
\end{align}
This equation suggests a novel way to analyze $c_V$
through the measurement of the energy fluctuation, which becomes 
possible with the correctly normalized EMT operator.
More interesting quantities related to the EMT correlators are 
transport coefficients.
Shear viscosity, for example, is related to the correlation
function of $T_{12}$ via the Kubo formula as \cite{Kubo}
\begin{align}
  \eta = \frac1{TV} \int_0^\infty dt 
  \langle \bar{T}_{12}(t); \bar{T}_{12}(0) \rangle,
  \label{eq:eta}
\end{align}
where $\langle O_1; O_2 \rangle$ is the Kubo's canonical
correlation and $\bar{O} = \int_V d^3 x O(x)$.
The transport coefficients are important quantities to 
understand dynamical time evolution of relativistic heavy ion 
collisions \cite{Romatschke:2009im}.
The measurement of transport coefficients with Kubo formula, 
however, requires real-time correlation function of the EMT, 
while the lattice simulations 
can obtain only imaginary time correlators directly 
\cite{Nakamura:2004sy,Meyer:2011gj}.
The analytic continuation from imaginary-time
functions to real time is needed, which, however, is a 
nontrivial procedure requiring an accurate measurement of 
the EMT correlator.

The application of the EMT operator is not limited to the 
physics of nonzero temperature.
For example, with the correctly normalized EMT operator one 
would be able to measure the energy and momentum distributions 
in hadrons and flux tube \cite{Bali:1994de} directly on the lattice.
The correlation functions of the EMT in vacuum are also
a useful probe to understand vacuum structure of the gauge theory.

Recently, a construction of the EMT in gauge theory 
is proposed in Ref.~\cite{Suzuki:2013gza} on the basis of 
the Yang-Mills gradient flow \cite{Luscher:2010iy} and the 
so-called small flow time expansion \cite{Luscher:2011bx}.
In this method, the constructed EMT is renormalized and does
not have divergences, and this property is kept even for the
lattice regularization.
Moreover, the statistical error is expected to be reduced
thanks to the cooling nature of the gradient flow.
Therefore, this method can become a solution for the
longstanding problem on the measurement of the EMT on the
lattice.
In this manuscript, we report the first numerical analysis of
the EMT in this method in lattice simulations \cite{FlowQCD}.

\section{Gradient flow}
\label{sec:GF}

Because we use the Yang-Mills gradient flow to define the EMT,
let us first give a brief review on this concept.
The gradient flow for the Yang-Mills gauge field is the 
continuous transformation of the field defined by the 
differential equation \cite{Luscher:2010iy}
\begin{align}
\frac{d A_\mu}{dt} = - g_0^2 
\frac{ \partial S_{\rm YM}(t)}{ \partial A_\mu }
= D_\nu G_{\nu\mu} ,
\label{eq:GF}
\end{align}
with the Yang-Mills action $S_{\rm YM}(t)$ composed of $A_\mu(t)$.
Color indices are suppressed for simplicity.
The initial condition at $t=0$ is taken for the field in the
conventional gauge theory; $A_\mu(0)=A_\mu$.
The flow time $t$, which has a dimension of inverse mass squared, 
is a parameter which controls the transformation.
The gauge field is transformed along the steepest descent 
direction as $t$ increases.
At the tree level, Eq.~(\ref{eq:GF}) is rewritten as 
\begin{align}
  \frac{d A_\mu}{dt} = \partial_\nu \partial_\nu A_\mu
  + {\rm (gauge ~ dependent ~ terms)}.
\label{eq:diffusion}
\end{align}
Neglecting the gauge dependent terms, Eq.~(\ref{eq:diffusion})
is the diffusion equation in four-dimensional space.
For positive $t$, therefore, the gradient flow acts as 
the cooling of the gauge field with the smearing radius 
$\sqrt{8t}$.
In Ref.~\cite{Luscher:2011bx}, it is rigorously proved that all 
composite operators composed of $A_\mu(t)$ take finite values for $t>0$.
This property ensures that observables at $t>0$ are 
automatically renormalized.

Recently, the gradient flow has been applied for various purposes,
such as the scale setting, analysis of the running coupling, 
measurement of topology, as well as the studies on their properties 
on the lattice; see reviews in Refs.~\cite{Luscher:2013vga,Ramos}
and references \cite{Luscher:2011bx,Suzuki:2013gza,DelDebbio:2013zaa,
Luscher:2013cpa,Makino:2014taa,Fodor:2014cpa}.
In the present study, we use the gradient flow to define 
the EMT operator using the small flow time expansion 
\cite{Luscher:2011bx,Suzuki:2013gza,DelDebbio:2013zaa}.

\section{Small flow time expansion and the EMT}
\label{sec:SFTE}

In the present study, we consider the EMT defined by 
the small flow time expansion \cite{Suzuki:2013gza}.
In order to illustrate the concept of small flow time expansion, 
let us consider a composite operator $\tilde{O}(t,x)$ 
composed of the field $A_\mu(t)$ at positive flow time.
The small flow time expansion asserts that in the small $t$ 
limit this operator can be written by a superposition of 
operators of the original gauge theory at $t=0$ as 
\begin{align}
  \tilde{O}(t,x) \xrightarrow[t\to0]{} \sum_i c_i(t) O_i^{\rm R}(x) ,
  \label{eq:SFTE}
\end{align}
where $O_i^{\rm R}(x)$ on the right-hand side represents renormalized 
operators in some regularization scheme in the original gauge 
theory at $t=0$ with the subscript $i$ denoting different operators,
and $x$ represents the coordinate in four dimensional space-time.

The relation Eq.~(\ref{eq:SFTE}) is reminiscent of the 
operator product expansion (OPE).
While a product of operators at different space-time 
points are expanded by local operators 
in the OPE, Eq.~(\ref{eq:SFTE}) relates
an operator at $t>0$ with those of $t=0$.
The validity of this expansion is expected from the 
cooling nature of the gradient flow; since the operator 
$\tilde{O}(t,x)$ depends on the fundamental gauge theory 
only inside the smearing radius $\sqrt{8t}$ around $x$, 
$\tilde{O}(t,x)$ should be described by the local property 
of the fundamental theory at $x$ in $t\to0$ limit.
Similarly to the Wilson coefficients in the OPE, 
the coefficients $c_i(t)$ in Eq.~(\ref{eq:SFTE}) 
can be calculated perturbatively \cite{Luscher:2011bx,Suzuki:2013gza}.

Using Eq.~(\ref{eq:SFTE}), one can define the renormalized 
EMT \cite{Suzuki:2013gza}.
For this purpose, we first consider dimension-four
gauge-invariant operators for operators of the 
left-hand side in Eq.~(\ref{eq:SFTE}).
In pure gauge theory, there are two such operators;
\begin{align}
  U_{\mu\nu}(t,x) &= G_{\mu\rho}^a (t,x)G_{\nu\rho}^a (t,x)
  -\frac14 \delta_{\mu\nu}G_{\rho\sigma}^a(t,x)G_{\rho\sigma}^a(t,x), 
  \label{eq:U}
  \\
  E(t,x) &= \frac14 G_{\mu\nu}^a(t,x)G_{\mu\nu}^a(t,x).
  \label{eq:E}
\end{align}
Although Eqs.~(\ref{eq:U}) and (\ref{eq:E}) are quite similar 
to the traceless-part and trace of the EMT, they are not 
the EMT since $G_{\mu\nu}(t,x)$ is defined at nonzero 
flow time $t>0$.
Because these operators are gauge invariant,
when they are expanded as in Eq.~(\ref{eq:SFTE})
only gauge invariant operators can appear in the 
right-hand side.
Such operator with the lowest dimension is an identity operator.
In the expansion of the traceless operator Eq.~(\ref{eq:U}), 
however, the constant term cannot appear.
The next gauge-invariant operators are the dimension-four EMTs.
Up to this order, therefore, the small flow time expansions
of Eqs.~(\ref{eq:U}) and (\ref{eq:E}) are given by 
\begin{align}
   U_{\mu\nu}(t,x)
   &=\alpha_U(t)\left[
   T_{\mu\nu}^R(x)-\frac14 \delta_{\mu\nu}T_{\rho\rho}^R(x)\right]
   +O(t),
\label{eq:(2)}\\
   E(t,x)
   &=\left\langle E(t,x)\right\rangle_0
   +\alpha_E(t)T_{\rho\rho}^R(x)
   +O(t),
\label{eq:(3)}
\end{align}
where $\langle\cdot\rangle_0$ is vacuum expectation value and 
$T_{\mu\nu}^R(x)$ is the correctly normalized conserved EMT 
with its vacuum expectation value subtracted. Abbreviated are 
the contributions from the operators of dimension~$6$ or higher, 
which are proportional to powers of $t$ because of dimensional 
reasons, and thus suppressed for small $t$.

Combining relations Eqs.~(\ref{eq:(2)}) and~(\ref{eq:(3)}), we have
\begin{align}
   T_{\mu\nu}^R(x)
   =\lim_{t\to0}\left\{\frac{1}{\alpha_U(t)}U_{\mu\nu}(t,x)
   +\frac{\delta_{\mu\nu}}{4\alpha_E(t)}
   \left[E(t,x)-\left\langle E(t,x)\right\rangle_0 \right]\right\}.
\label{eq:T^R}
\end{align}
The coefficients $\alpha_U(t)$ and $\alpha_E(t)$ are calculated
perturbatively in Ref.~\cite{Suzuki:2013gza} as
\begin{align}
   \alpha_U(t)
   &=\Bar{g}(1/\sqrt{8t})^2
   \left[1+2b_0\Bar{s}_1\Bar{g}(1/\sqrt{8t})^2+O(\Bar{g}^4)\right],
\label{eq:(5)}
\\
   \alpha_E(t)
   &=\frac{1}{2b_0}\left[1+2b_0\Bar{s}_2
   \Bar{g}(1/\sqrt{8t})^2+O(\Bar{g}^4)\right].
\label{eq:(6)}
\end{align}
Here $\Bar{g}(q)$ denotes the running gauge coupling in the
$\overline{\text{MS}}$ scheme with the choice, $q=1/\sqrt{8t}$, and
\begin{align}
\Bar{s}_1 &= \frac{7}{16}+\frac{1}{2}\gamma_E-\ln2\simeq0.032960651891,
\\
\Bar{s}_2 &= \frac{109}{176}-\frac{b_1}{2b_0^2}
=\frac{383}{1936}\simeq0.19783057851,
\end{align}
with 
$b_0=\frac{1}{(4\pi)^2}\frac{11}{3}N_c$,
$b_1=\frac{1}{(4\pi)^4}\frac{34}{3}N_c^2$,
and~$N_c=3$. Note that a non-perturbative determination of~$\alpha_{U,E}(t)$ is
also proposed recently~\cite{DelDebbio:2013zaa,Patella:2014dsa}.

There are two important
observations: (i) The right-hand side of Eq.~(\ref{eq:T^R}) is 
independent of the regularization because of its UV finiteness,
so that one can take, e.g., the lattice regularization scheme;
(ii) since flowed fields at $t>0$ depend on the fundamental
fields at $t=0$ in the space-time region of radius $≃\sqrt{8t}$, 
the statistical noise in calculating the right-hand side of 
Eq.~(\ref{eq:T^R}) is suppressed for finite $t$.

\section{Numerical procedure on the lattice}
\label{sec:num}

The formula Eq.~(\ref{eq:T^R}) indicates that $T^R_{\mu\nu}(x)$
can be obtained by the small $t$ limit of the gauge-invariant 
local operators defined through the gradient flow. 
To perform the numerical analysis on the lattice, 
we take the following steps:
\begin{enumerate}
\item
Generate gauge configurations at $t=0$ on a space-time lattice
with a standard algorithm with the lattice spacing $a$ and 
the lattice size $N_s^3\times N_\tau$.
\item
Solve the gradient flow for each configuration to obtain the
flowed link variables for $t>0$.
\item
Construct $U_{\mu\nu}(t,x)$ and~$E(t,x)$
in terms of the flowed link variables
at each $t$.
\item
Determine the right hand side of Eq.~(\ref{eq:T^R}) for 
each $t$, and measure expectation values and correlation
functions of the EMT by regarding the right 
hand side of Eq.~(\ref{eq:T^R}) as the EMT.
\item
Carry out an extrapolation toward~$(a,t)=(0,0)$.
\end{enumerate}

Note that this analysis has to be performed 
in the fiducial window, $a\ll\sqrt{8t}\ll R$. Here,
$R$~is an infrared cutoff scale such as~$\Lambda_{\text{QCD}}^{-1}$
or~$T^{-1}=N_\tau a$ at which the perturbative analysis used in 
Eq.~(\ref{eq:T^R}) is violated. 
The first inequality is necessary to suppress 
finite $a$ corrections, while the second one is required
to suppress non-perturbative corrections and finite volume
effects.
We will later see that the upper limit of the fiducial window 
$R$ is given by by $T^{-1}=N_\tau a$ for temperatures 
analyzed in this study.

\section{Numerical setup}
\label{sec:setup}

To demonstrate that Eq.~(\ref{eq:T^R}) 
can successfully be used in the measurement of the EMT on the 
lattice, 
we consider the $SU(3)$ gauge theory defined on a four-dimensional 
Euclidean lattice, whose thermodynamics has been extensively 
studied by the integral method
\cite{Boyd:1996bx,Okamoto:1999hi,Umeda:2008bd,Borsanyi:2012ve}. 
We consider the Wilson plaquette gauge action under the periodic
boundary condition with several different 
$\beta=6/g_0^2$ with $g_0$ being the bare coupling constant.
The gradient flow in the $t$-direction is numerically solved by 
the Runge--Kutta method.
We have checked that the accumulation errors due to the Runge--Kutta
method is more than two order smaller than the statistical errors
in all analyses.

\begin{table}[t]
\begin{center}
\begin{tabular}{|c||c|c|c||c|}
\hline
$N_\tau$ & 6 & 8 & 10 & $T/T_c$ \\
\hline
         & 6.20 & 6.40 & 6.56 &  1.65 \\ 
$\beta$  & 6.02 & 6.20 & 6.36 &  1.24 \\
         & 5.89 & 6.06 & 6.20 &  0.99 \\
\hline
\end{tabular}
\caption{
Values of $\beta$ and~$N_\tau$ for each temperature for Setting~1 
\cite{FlowQCD}.}
\label{table:Nt_beta1}
\end{center}
\end{table}

\begin{table}[t]
\begin{center}
\begin{tabular}{|c||c|c|c|c|c|}
\hline
$N_\tau$ & $10$    & $12$    & $14$   &   $16$ & $20$ \\
\hline
$N_s$    & $64$    & $64$    & $64$   &   $64$ & $128$ \\
\hline
$\beta$ & $6.829$ & $6.971$ & $7.093$ & $7.200$ & $7.376$\\ 
\hline
\end{tabular}
\caption{
Values of $N_\tau$, $N_s$ and $\beta$ for $T/T_c=2.31$ for Setting~2.}
\label{table:Nt_beta2}
\end{center}
\end{table}

The numerical analyses reported in the following have been 
performed with the following two settings.
\begin{itemize}
\item
{\bf Setting~1:} 
The first analysis is performed 
on~$N_s^3\times N_\tau=32^3\times(6,8,10)$ lattices \cite{FlowQCD}.
Gauge configurations are generated by the pseudo-heatbath algorithm with the
over-relaxation, mixed in the ratio of~$1:5$. We call one pseudo-heatbath
update sweep plus five over-relaxation sweeps as a ``Sweep''. To eliminate the
autocorrelation, we take $200$--$500$ Sweeps between measurements. The number
of gauge configurations for the measurements at finite~$T$ is~$300$.
Statistical errors are estimated by the jackknife method.
To relate $T/T_c$ and corresponding $\beta$ for each~$N_\tau$, we first use the
relation between $a/r_0$ ($r_0$ is the Sommer scale) and~$\beta$ given by the
ALPHA Collaboration~\cite{Guagnelli:1998ud}. The resultant values of
$Tr_0=[N_\tau(a/r_0)]^{-1}$ are then converted to~$T/T_c$ by using the result
at~$\beta=6.20$ in~Ref.~\cite{Boyd:1996bx}. Nine combinations
of~$(N_\tau,\beta)$ and corresponding $T/T_c$ obtained by this procedure are
shown in~Table~\ref{table:Nt_beta1}.
\item
{\bf Setting~2:}
The second analysis is performed on finer lattices with $N_s=64$ 
and $128$ with several values of $N_\tau$.
The lattice spacing of each $\beta$ are determined originally \cite{scale}
by the measurements of $t_0$ \cite{Luscher:2010iy} and $w_0$ 
\cite{Borsanyi:2012zs} with the gradient flow, and 
the simulation is performed for several values of $T/T_c$.
In this manuscript, we present the result only for $T/T_c=2.31$.
The parameters $\beta$, $N_\tau$ and $N_s$ for this $T$ are 
shown in Table~\ref{table:Nt_beta2}.
To eliminate the autocorrelation, we take $200$ Sweeps 
between configurations.
The estimate of the statistical error by the jackknife method
suggests that the autocorrelation between each configuration
is well suppressed.
The typical number of gauge configurations 
for each measurement at finite $T$ is $2000$.
\end{itemize}

To analyze the EMT with Eq.~(\ref{eq:T^R}), we measure
$G^a_{\mu\nu}(t,x)$ written in terms of the clover leaf
representation on the lattice. To subtract out the $T=0$ contribution,
$\langle E(t,x)\rangle_0$, we carry out simulations on a $N_\tau=N_s$ 
lattice for each $\beta$. Note that this vacuum subtraction is
required only for the measurement of the trace anomaly
$\Delta=\varepsilon-3P$, while no subtraction is needed for the 
measurement of the entropy density~$s=(\varepsilon+P)/T$.
For~$\Bar{g}$ in~$\alpha_U(t)$ and $\alpha_E(t)$ given by~Eqs.~(\ref{eq:(5)})
and~(\ref{eq:(6)}), we use the four-loop running coupling with the scale
parameter determined by the ALPHA Collaboration,
$\Lambda_{\overline{\text{MS}}}=0.602(48)/r_0$~\cite{Capitani:1998mq}. 
In the following, we present the numerical results of the 
thermodynamics for two Settings in Secs.~\ref{sec:thermo1} and 
\ref{sec:thermo2}, respectively, and then discuss 
the correlation function of the EMT in Setting 2 
in Sec.~\ref{sec:corr}.

\section{Thermodynamics: Setting 1}
\label{sec:thermo1}

\begin{figure}[t]
\begin{center}
\includegraphics[scale=0.45]{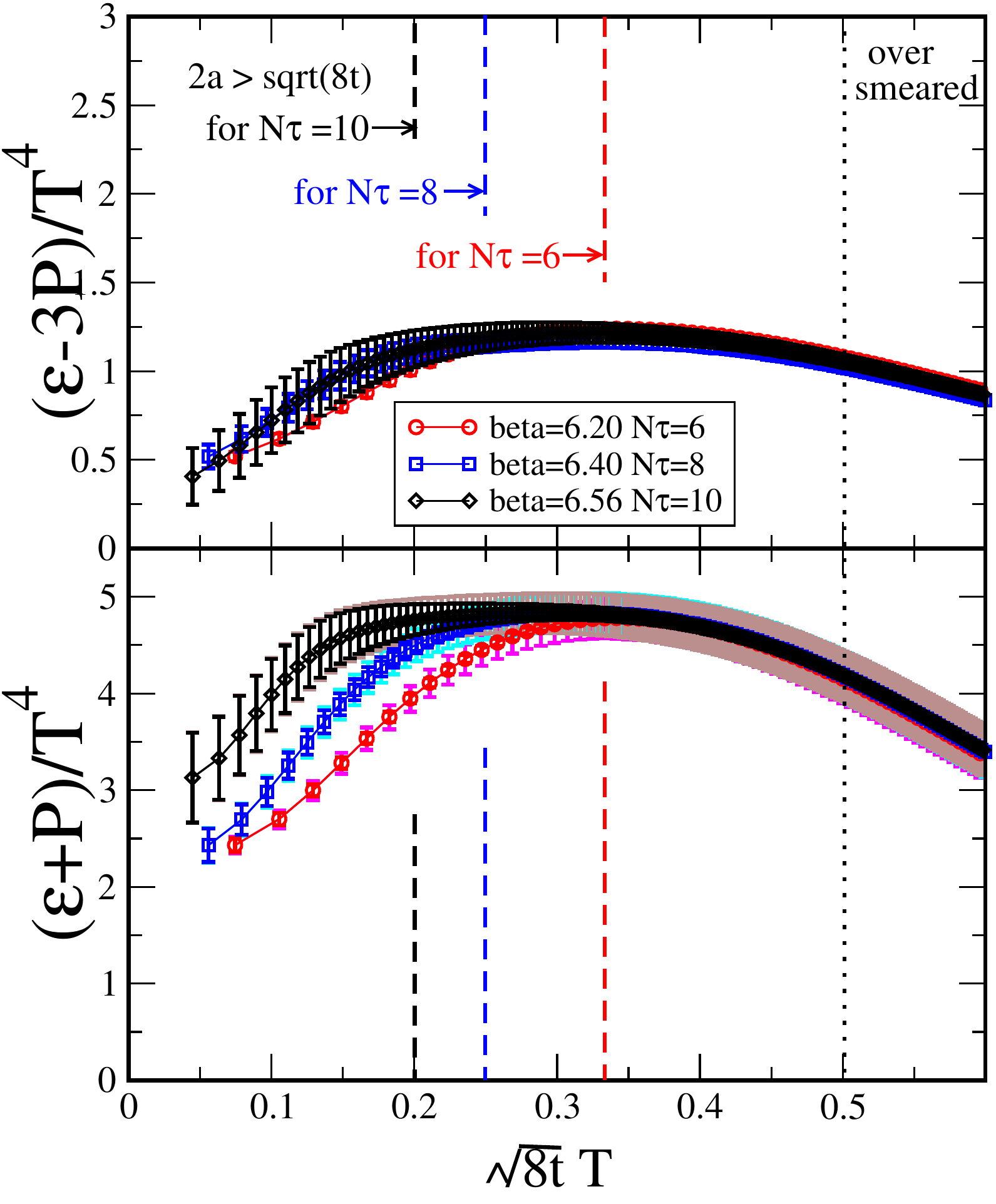}
\caption{Flow time dependence of the dimensionless interaction measure (top
panel) and the dimensionless entropy density (bottom panel) for different
lattice spacings at fixed~$T/T_c=1.65$ \cite{FlowQCD}. 
The circles (red) the squares (blue),
and the diamonds (black) correspond to~$N_\tau=6$, $8$, and~$10$, respectively.
The bold error bars denote the statistical errors, while the thin error bars
(brown, cyan, and magenta) include both statistical and systematic errors.}
\label{fig:raw-data-1.65Tc}
\end{center}
\end{figure}

In this section, we first present the numerical results on bulk
thermodynamics obtained by the Setting~1 \cite{FlowQCD}.
In Fig.~\ref{fig:raw-data-1.65Tc}, we show the 
results for the dimensionless trace anomaly 
$\Delta/T^4=(\varepsilon-3P)/T^4$ and the dimensionless entropy 
density $s/T^3=(\varepsilon+P)/T^4$ at~$T=1.65T_c$ as a
function of the dimensionless flow parameter~$\sqrt{8t}T$ \cite{FlowQCD}. 
The bold bars denote the statistical errors, while the thin 
(light color) bars show the statistical and systematic errors 
including the uncertainty of~$\Lambda_{\overline{\text{MS}}}$. 
We found that the statistical error is
dominant in the small $t$ region for both $\Delta/T^4$ and~$s/T^3$, while the
systematic error originating from the scale parameter becomes significant
for~$s/T^3$ in the large $t$ region.

The fiducial window discussed in Sec.~\ref{sec:num} is indicated by the dashed
lines in~Fig.~\ref{fig:raw-data-1.65Tc}. The lower limit of the window, beyond
which the lattice discretization error grows, is set to
be~$\sqrt{8t_{\rm min}}=2a$, where we take into account the fact that our clover
leaf operator extends the size~$2a$. The upper limit of the window, beyond
which the smearing by the gradient flow exceeds the temporal lattice size, is
set to be $\sqrt{8t_{\rm max}}=1/(2T)=N_\tau a/2$.

The data in~Fig.~\ref{fig:raw-data-1.65Tc} show, within the error bars, that
(i)~the plateau appears inside the preset fiducial window
($2/N_\tau<\sqrt{8t}T<1/2$) for each~$N_\tau$, and (ii)~the plateau extends to
the smaller~$t$ region as $N_\tau$ increases or equivalently as $a$ decreases.
It should be remarked that only $300$ gauge configurations are necessary to
obtain these results. Similar plateaues as in~Fig.~\ref{fig:raw-data-1.65Tc}
also appear inside the fiducial window for other temperatures, $T/T_c= 1.24$
and~$0.99$, with comparable error bars. These features imply that 
the double extrapolation $(a,t)\to(0,0)$ is indeed
possible.

\begin{figure}[t]
\begin{center}
\includegraphics[width=0.49\textwidth]{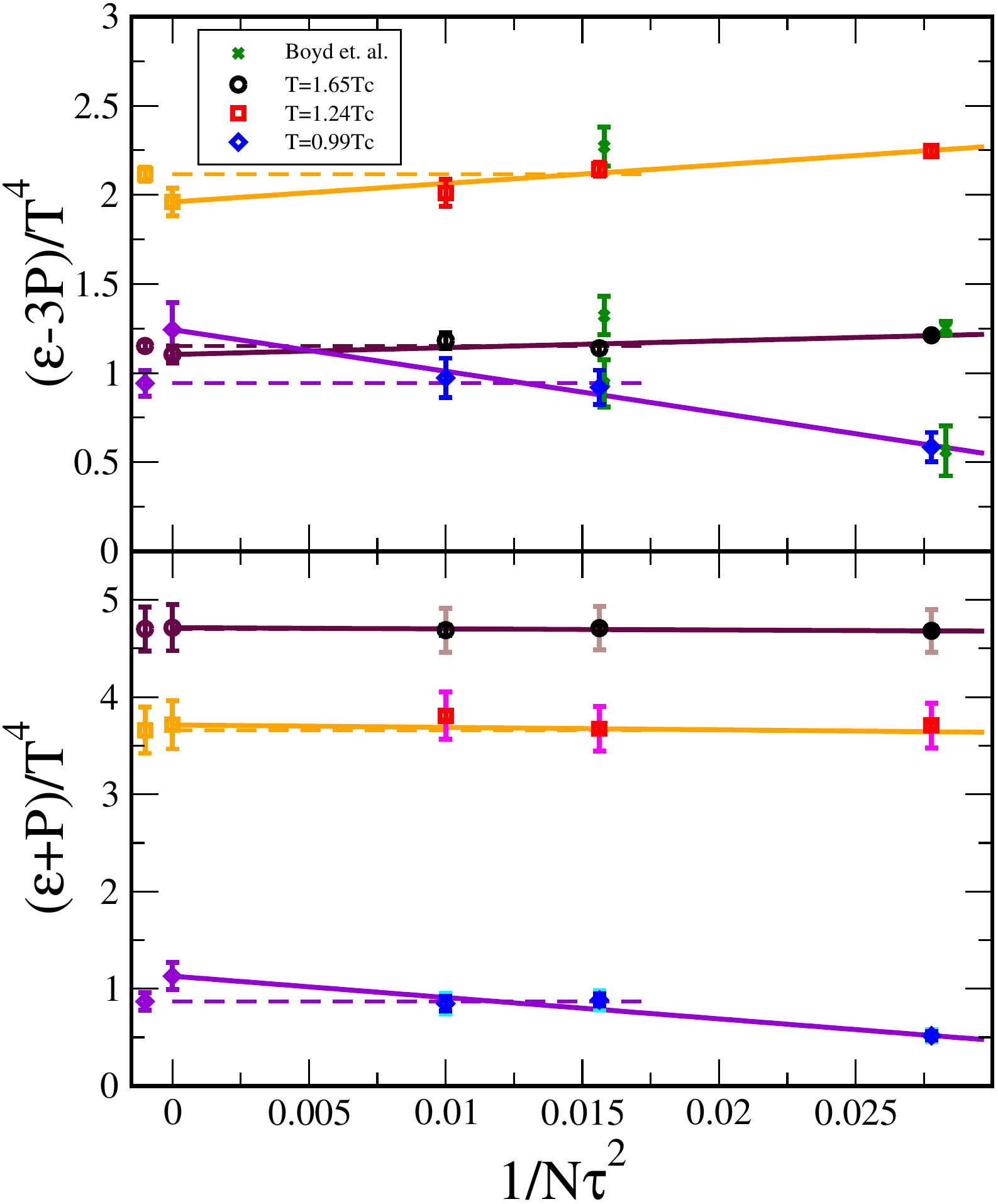}
\includegraphics[width=0.49\textwidth]{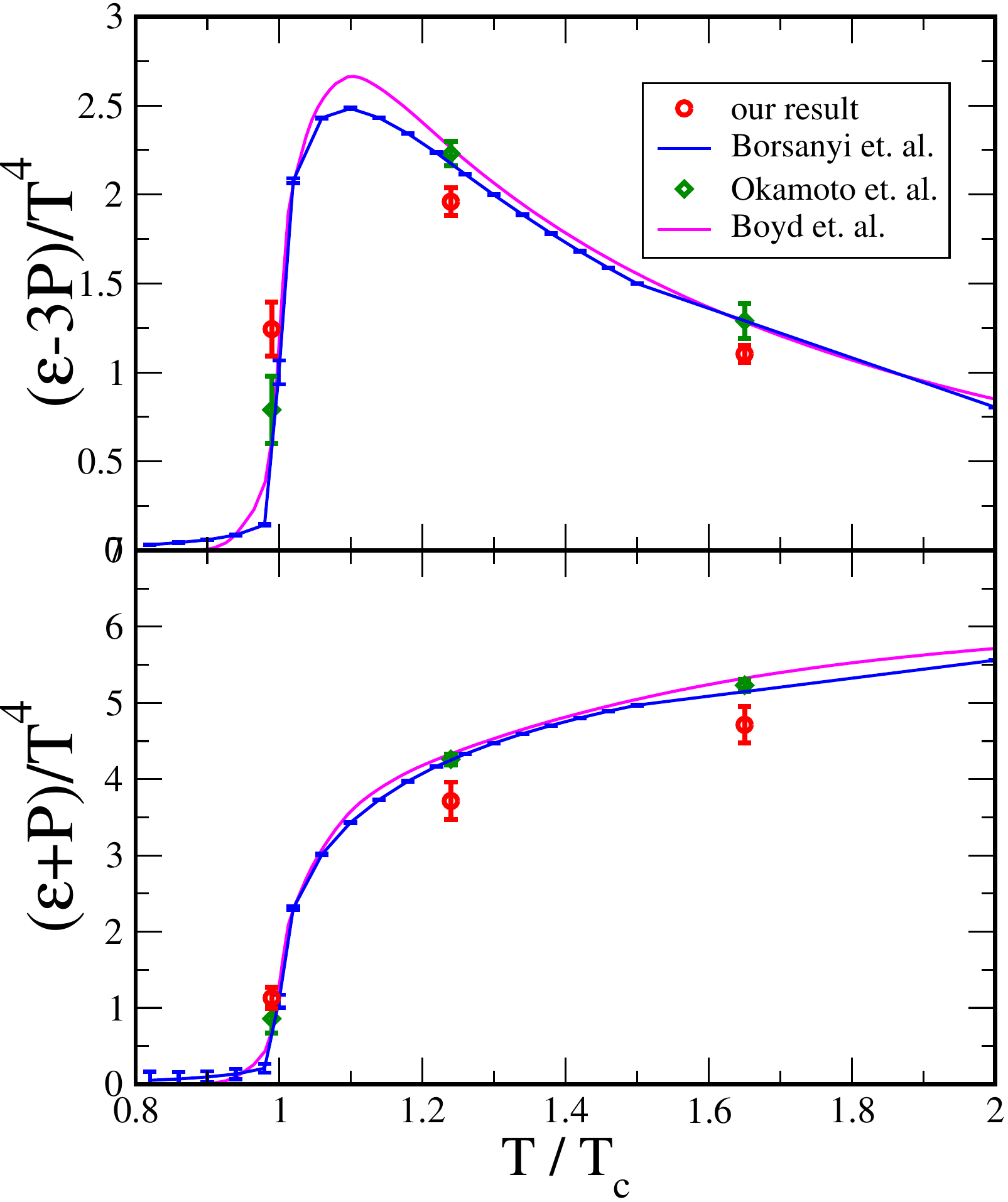}
\caption{
{\bf Left:} Continuum extrapolation of the thermodynamic quantities
for~$T/T_c=1.65$, $1.24$, and~$0.99$. Solid lines and dashed lines correspond
to the three-point linear fit and two-point constant fit as a function
of~$1/N_\tau^2$, respectively. Extrapolated values of the former (latter) are
shown at~$1/N_\tau^2=0$ ($1/N_\tau^2=-0.001$). The cross symbols in the top
panel are the data of~Ref.~\cite{Boyd:1996bx} with the same lattice setup.
{\bf Right:}
Continuum limit of the interaction measure and entropy density
obtained by the gradient flow for~$T/T_c=1.65$, $1.24$, and~$0.99$ with $300$
gauge configurations. 
The results obtained by the integral method in Refs.~\cite{Boyd:1996bx}, 
\cite{Okamoto:1999hi} and \cite{Borsanyi:2012ve} are also plotted.
}
\label{fig:cont-lim}
\end{center}
\end{figure}

Our lattice results at fixed~$T$ with three different lattice spacings allow us
to take the continuum limit. First, we pick up a flow time $\sqrt{8t}T=0.40$
which is in the middle of the fiducial window. Then we extract $\Delta/T^4$
and~$s/T^3$ for each set of~$N_\tau$ and~$\beta$. 
In the left panel of Fig.~\ref{fig:cont-lim}, resultant values
taking into account the statistical errors (bold error bars) and the
statistical plus systematic errors (thin error bars) are shown. The lattice
data for~$\Delta/T^4$ with the same lattice setup at~$N_\tau=6$ and~$8$
in~Ref.~\cite{Boyd:1996bx} are also shown by the cross (green) symbols in the
top panel; our results with $300$ gauge configurations have substantially
smaller error bars at these points.

\begin{figure}[t]
\begin{center}
\includegraphics[width=0.7\textwidth]{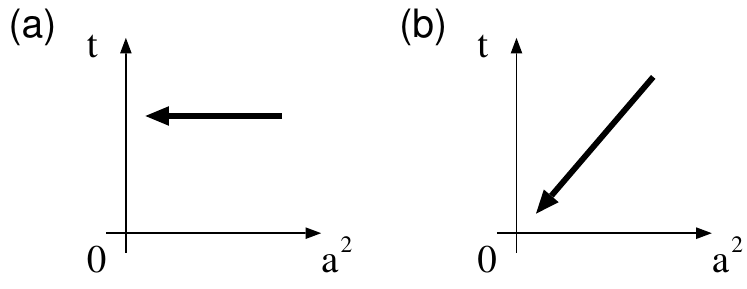}
\caption{
Continuum extrapolation in Setting~1 (a) and Setting~2 (b).
}
\label{fig:extrapol}
\end{center}
\end{figure}

We note that the continuum extrapolation in this analysis \cite{FlowQCD}
is taken with fixed $t$; this procedure is 
graphically shown in Fig.~\ref{fig:extrapol} (a).
Strictly speaking, however, the double limit $(a,t)\to(0,0)$ has 
to be taken as mentioned above.
Although we have checked that different choices of $t$ do not change 
the final results within the error bar as long as it is in the plateau 
region in the analysis in Setting~1, in the next section we will see that 
the limit with fixed $t$ shows a deviation compared with the double
limit when the statistics is improved.

The horizontal axis of the left panel of Fig.~\ref{fig:cont-lim}, 
$1/N_\tau^2$, is a variable
suited for making continuum extrapolation of the thermodynamic
quantities~\cite{Boyd:1996bx}. We consider two ways of extrapolation: A linear
fit with the data at~$N_\tau=6$, $8$, and~$10$ (the solid lines
in the left panel of Fig.~\ref{fig:cont-lim}), and a constant fit 
with the data at~$N_\tau=8$
and~$10$ (the dashed lines in the left panel of Fig.~\ref{fig:cont-lim}). 
In both fits, the
correlation between the errors due to the common systematic error
from~$\Lambda_{\overline{\text{MS}}}$ is taken into account. The former fit is
used to determine the central value in the continuum limit whose error is
within~$\pm12\%$ even at our lowest temperature. The latter is used to estimate
the systematic error from the scaling violation whose typical size
is~$\pm4\%$ at high temperature and~$\pm24\%$ at low temperature.

We have analyzed various systematic errors; the perturbative expansion
of~$\alpha_{U,E}(t)$, the running coupling~$\Bar{g}$, the scale parameter, and
the continuum extrapolation. We found that the dominant errors in the present
lattice setup are those from~$\Lambda_{\overline{\text{MS}}}$ and the continuum
extrapolation, which are included in the left panel of Fig.~\ref{fig:cont-lim}. 
To reduce these
systematic errors, finer lattices are quite helpful: They make the plateau
in~$\sqrt{8t}T$ wider by reducing the lower limit of the fiducial window, so
that the continuum extrapolation becomes easier. Also, larger aspect ratio
$N_s/N_\tau$ would be helpful to guarantee the thermodynamic limit. 


Finally, we plot, in the right panel of Fig.~\ref{fig:cont-lim}, 
the continuum limit
of~$\Delta/T^4$ and~$s/T^3$ obtained by the linear fit of the $N_\tau=6$, $8$,
and~$10$ data (the solid lines) in the left panel of Fig.~\ref{fig:cont-lim} for $T/T_c=1.65$,
$1.24$, and~$0.99$. 
For comparison to the existing data of the $SU(3)$ gauge theory after
the continuum extrapolation, the results of Refs.~\cite{Boyd:1996bx} 
\cite{Okamoto:1999hi} and \cite{Borsanyi:2012ve} obtained by the 
integral method are also plotted in the right panel of Fig.~\ref{fig:cont-lim}.
The results of the two different approaches are consistent
with each other within $2$~sigma level.
The figure, however, indicates that our results underestimate 
the observables compared with the previous ones.
This discrepancy in part come from the continuum extrapolation 
with fixed $t$. The correct double limit will be discussed 
in the next section.

\section{Thermodynamics: Setting~2}
\label{sec:thermo2}

\begin{figure}[t]
\begin{center}
\includegraphics[width=0.49\textwidth]{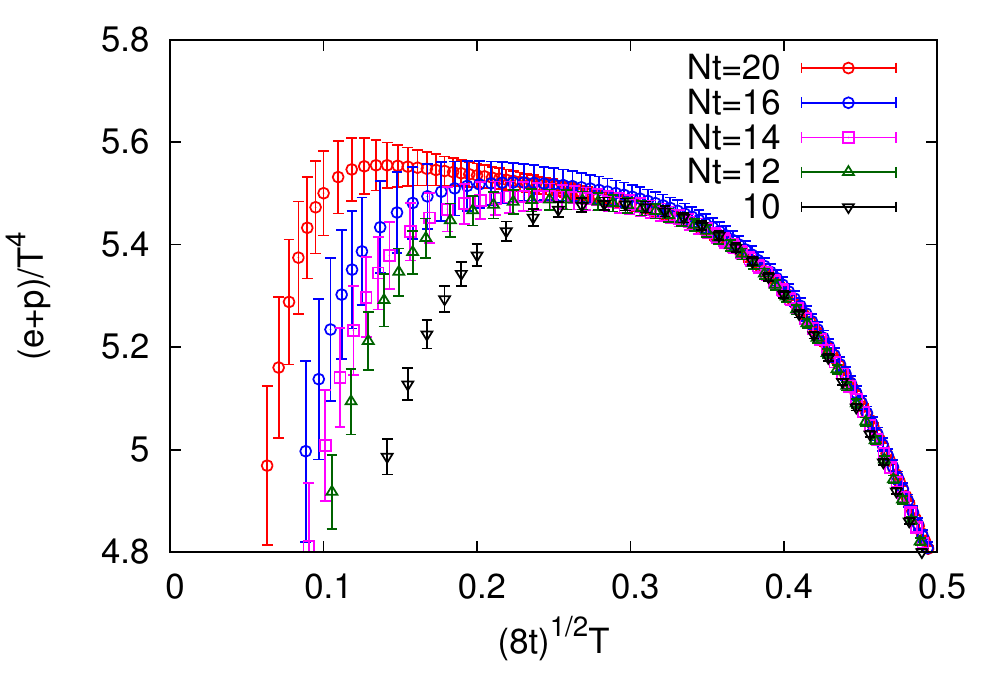}
\includegraphics[width=0.49\textwidth]{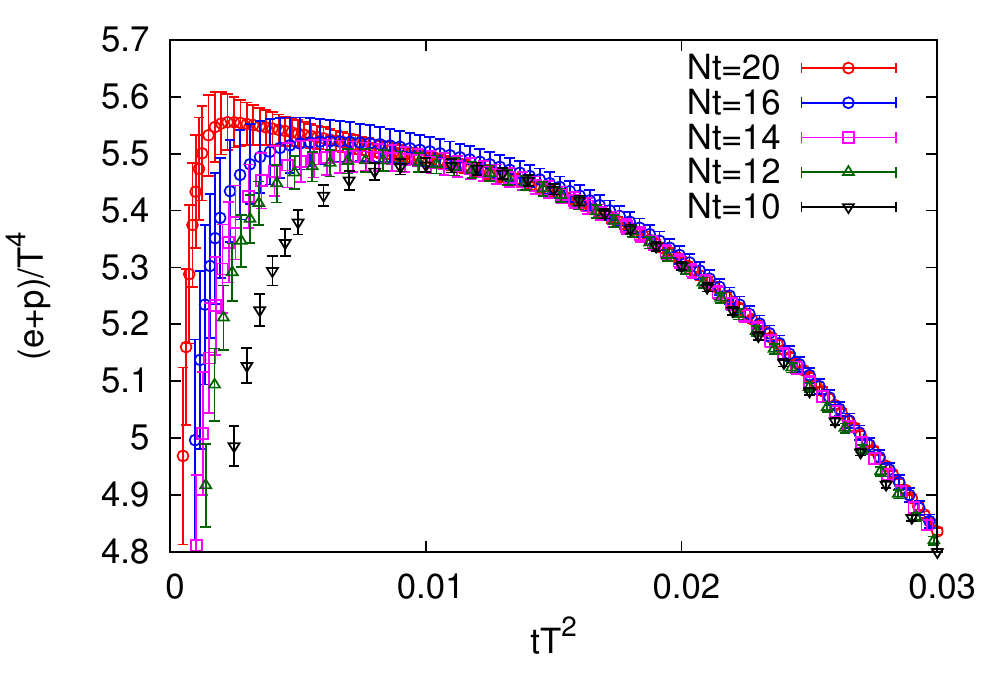}
\caption{
Flow time dependence of the dimensionless entropy density 
$s=(\varepsilon+P)/T$ at $T/T_c=2.31$ for several values of $N_\tau$ 
with $N_s=64$ and $128$.
The same result is plotted as functions of 
$\sqrt{8t}T$ (left panel) and $tT^2$ (right panel). 
The errorbars denote statistical error.
}
\label{fig:s231}
\end{center}
\end{figure}

Next, let us see the numerical results on thermodynamics  
with finer lattices with Setting~2.
In the left panel of Fig.~\ref{fig:s231}, we show the dimensionless
entropy density $s/T^3=(\varepsilon+P)/T^4$ as a function of 
$\sqrt{8t}T$ for $T/T_c=2.31$ with $N_s=64$ and $128$ 
for several values of $N_\tau$.
In this result, one can take a closer look at the 
$t$ dependence of $s/T^3$ thanks to the improved 
statistics and finer lattices.
In the previous section, we discussed that a plateau is observed
in the fiducial window, $2a<\sqrt{8t}<1/T$, in Setting~1.
Because the statistics is significantly improved in Setting~2, 
in Fig.~\ref{fig:s231} one sees a $t$ dependent structure 
in the fiducial window with $N_t=10$.
Next, while the $t$ dependence at small $t$ becomes more 
stable as the lattice spacing becomes finer, the $t$ dependence
does not show a plateau even on the finest lattice with 
$N_\tau=20$.
In the right panel of Fig.~\ref{fig:s231}, the same result is 
shown as a function of $tT^2$ instead of $\sqrt{8t}T$.
The figure indicates that $s/T^3$ behaves as a linear function 
of $t$ as the lattice spacing becomes finer.

The origin of the linear term can be understood as the 
contribution of higher dimensional operators in 
Eq.~(\ref{eq:T^R}) giving rise to $O(t)$ contribution 
to this formula.
In fact, the contribution of dimension six operators are 
proportional to $t$ in $t\to0$ limit.
Figure~\ref{fig:s231} shows that the effect of these terms
has nonnegligible contribution.
Such a contribution has to be subtracted 
to measure thermodynamic quantities.
A similar result is obtained for $\Delta/T^4$, while the $t$ 
dependence is weaker than the one in $s/T^3$.

The $O(t)$ effect discussed here can be eliminated 
by taking the double limit $(a,t)\to(0,0)$.
In order to take this double extrapolation, 
in this analysis we take $a\to0$ limit with fixed $t/a^2$; 
this procedure is graphically shown in Fig.~\ref{fig:extrapol} (b).
Here, the ratio $t/a^2$ has to be sufficiently large
so that $t$ and $a$ satisfy the condition $2a<\sqrt{8t}$,
or $t/a^2>1/2$.
Once this condition is satisfied, however, 
the double limit can be taken safely irrespective 
of the value of $t/a^2$.
In this extrapolation, it is expected that the 
discretization effect is of order $a^2$.
This is because the leading discretization error 
on thermodynamic observables is order $a^2$ \cite{Boyd:1996bx}, 
while the error arising from nonzero $t$ is 
of order $t$.

\begin{figure}[t]
\begin{center}
\includegraphics[width=0.49\textwidth]{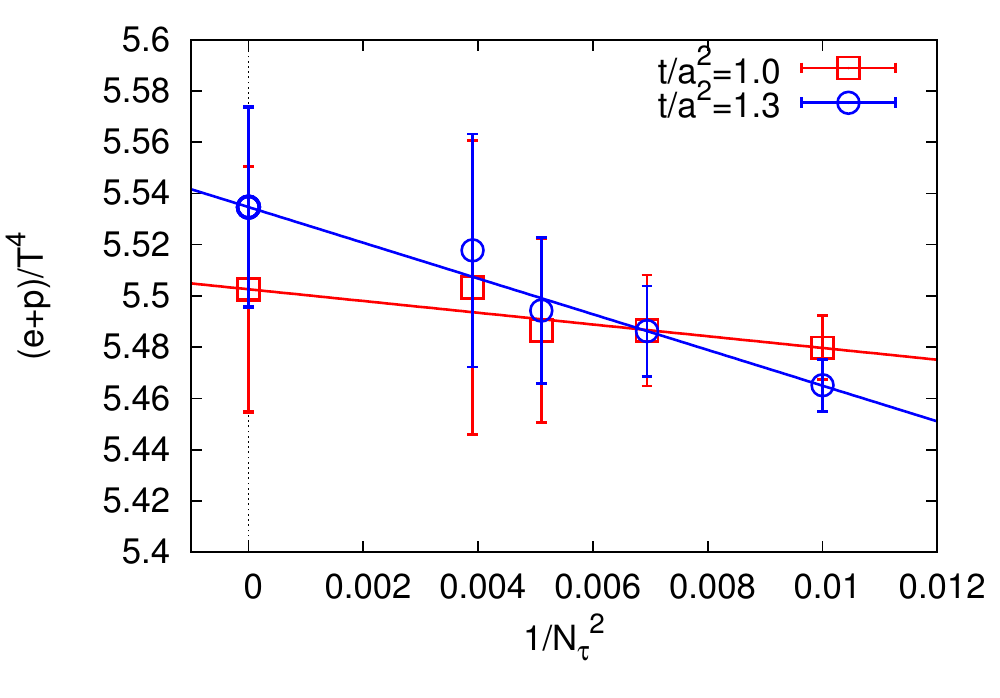}
\includegraphics[width=0.49\textwidth]{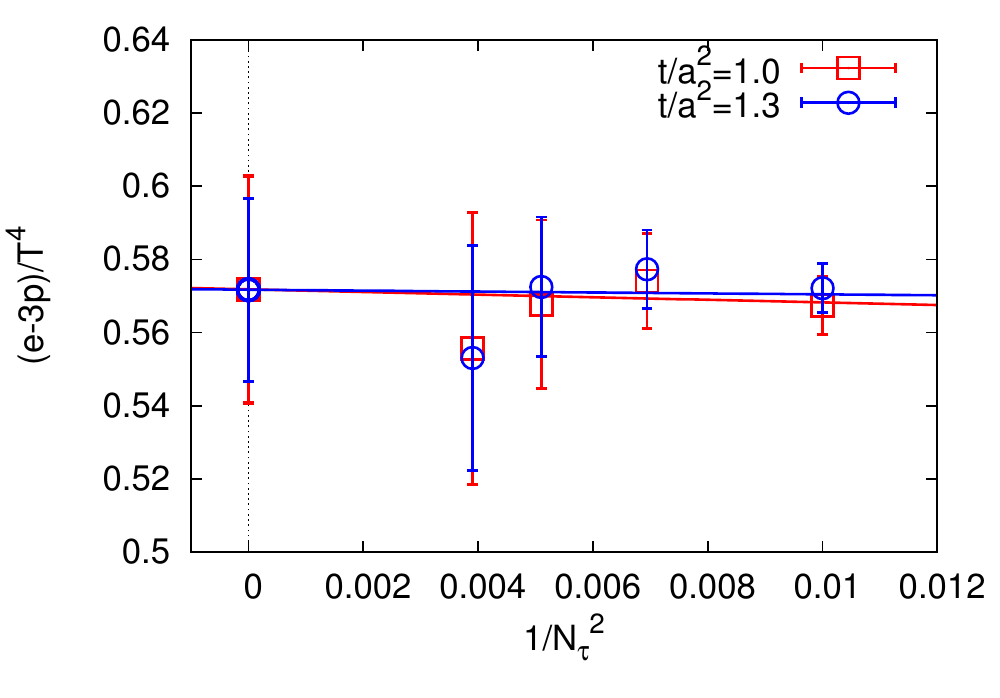}
\caption{
Continuum extrapolation of $s$ and $\Delta$.
Double extrapolation $(a,t)\to(0,0)$ is taken by 
taking $a\to0$ limit with fixed $t/a^2=1.0$ and $1.3$.
}
\label{fig:extrapolate231}
\end{center}
\end{figure}

In Fig.~\ref{fig:extrapolate231}, 
we show the result of the continuum extrapolation 
for $s/T^3$ and $\Delta/T^4$ with fixed $t/a^2$ 
for $t/a^2=1$ and $1.3$.
We choose $1/N_\tau^2$ for the horizontal axis and 
take a linear extrapolation because of the $a$ dependence
discussed above.
The result shows that the continuum extrapolation is 
stable and the extrapolations with different $t/a^2$ 
are consistent.
Because of the negative slope in Fig.~\ref{fig:s231},
the continuum extrapolation of $s/T^3$ obtained in this method 
gives slightly larger value compared with the result 
in the previous section.
The final results after the double extrapolation 
are consistent with the existing results in the integral 
method \cite{Boyd:1996bx,Borsanyi:2012ve}.

\section{Correlation function}
\label{sec:corr}

\begin{figure}[t]
\begin{center}
\includegraphics[width=0.6\textwidth]{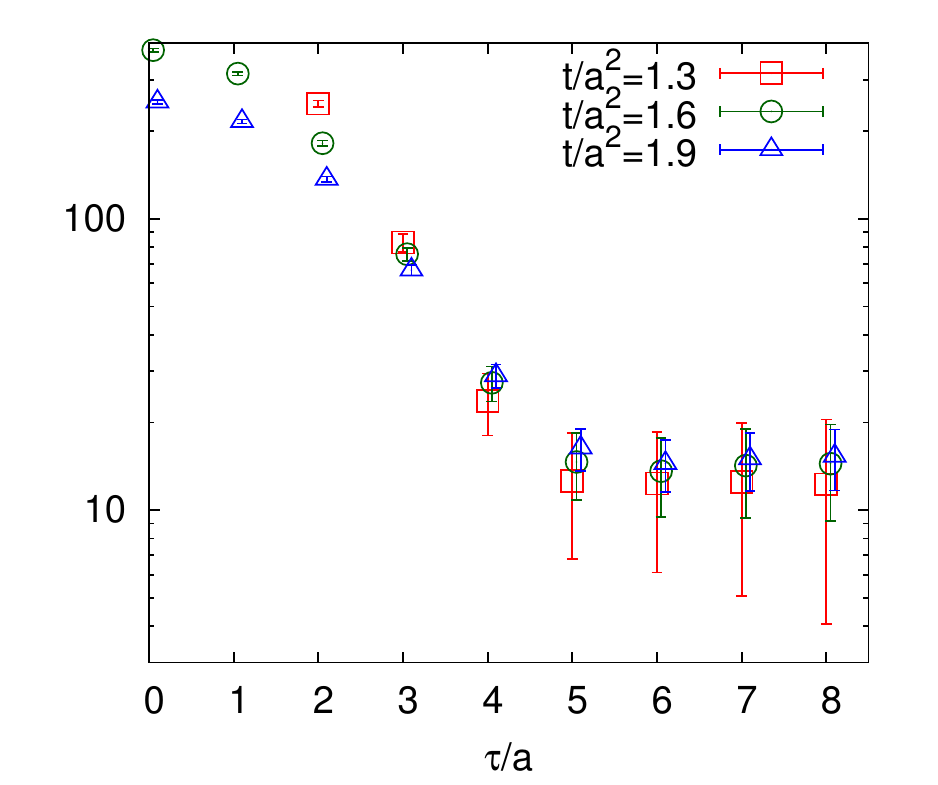}
\caption{
Imaginary time correlation function of the total energy, 
$C_E(\tau)/T^5 = \langle \delta \bar{T}_{00} (\tau) 
\delta \bar{T}_{00} (0) \rangle/(VT^5)$, 
for several values of $t/a^2$ for $T/T_c=2.31$ obtained on
$64^3\times16$ lattice.
}
\label{fig:correlator}
\end{center}
\end{figure}

Once the operator of the EMT on the lattice is obtained,
the application of this operator is not limited to the 
bulk thermodynamics which is given by expectation 
values of one point functions.
In this section, as an example of other applications 
we consider the correlation function of the EMT.
As discussed in Sec.~\ref{sec:intro}, 
the correlation functions of the EMT are important observables
related to transport properties of the medium and fluctuations.

In this manuscript, as an example of such analyses 
we present the numerical result on the Euclidean correlation 
function of the total energy,
\begin{align}
C_E(\tau) = \frac1V \langle \delta \bar{T}_{00} (\tau) 
\delta \bar{T}_{00} (0) \rangle,
\label{eq:corr}
\end{align}
with $\bar{T}_{00} = \int d^3 x T_{00}(x,\tau)$ and 
$\delta \bar{T}_{00} = \bar{T}_{00}-\langle\bar{T}_{00}\rangle$


When one investigates correlation functions of the EMT 
with Eq.~(\ref{eq:T^R}), one has to take care of 
the cooling nature of the gradient flow.
When correlation functions are observed on this field,
the smearing nature gives rise to artificial correlations 
in two point correlators when the two operators 
are not separated twice the smearing length.
Because of Eq.~(\ref{eq:diffusion}), the smearing radius
defined by the mean square distance in four dimensional space
is $\sqrt{8t}$ \cite{Luscher:2010iy}.
When one focuses on a smearing along one definite direction, 
on the other hand, the smearing length is $\sqrt{2t}$.
Therefore, the correlation function Eq.~(\ref{eq:corr})
receives an artificial effect due to the smearing 
for $\tau\lesssim2\sqrt{2t}$.

Keeping this precaution in mind, let us see the numerical 
result on Eq.~(\ref{eq:corr}).
In Fig.~\ref{fig:correlator}, we show $\tau$ dependence 
of the correlator Eq.~(\ref{eq:corr}) for several
values of the flow time $t$ for $T/T_c=2.31$ and $N_\tau=16$.
From the figure, one can see several interesting features.
First, $C_E(\tau)$ takes a large value for $\tau \lesssim 2\sqrt{2t}$
in which the above-mentioned smearing problem affects the 
correlation function.
The enhancement of $C_E(\tau)$ at $\tau \lesssim 2\sqrt{2t}$ has a 
clear $t$ dependence, which indicates that this enhancement is 
indeed the artificial behavior arising from the gradient flow.
On the other hand, the correlator for $\tau \gtrsim 2\sqrt{2t}$
does not have a statistically significant $t$ dependence.
This result implies that the correlation function can 
be obtained in this range.
Second, the errorbars of the result are more suppressed
for larger $t$.
For $t/a^2=1.9$, the error is less than $10\%$
with $2000$ configurations.
Compared with the previous studies on the EMT correlators
\cite{Nakamura:2004sy,Meyer:2011gj},
this number of statistics is significantly small.

Figure~\ref{fig:correlator} also shows that the 
correlator Eq.~(\ref{eq:corr}) takes a constant 
within statistics in the range $\tau \gtrsim 2\sqrt{2t}$ 
at which the smearing effect is well suppressed.
This result can be understood as a consequence of 
the energy conservation, because the energy conservation
requires 
\begin{align}
  \frac{\partial}{\partial\tau} C_E(\tau) 
  = \langle \frac{\partial}{\partial\tau} \bar{T}_{00} (\tau) 
  \bar{T}_{00} (0) \rangle
  = 0,
  \label{eq:d_tau}
\end{align}
for $\tau\ne0$.
The emergence of a plateau in Fig.~\ref{fig:correlator}, therefore, 
shows that the EMT defined in Eq.~(\ref{eq:T^R}) satisfies the 
energy conservation.

The constant of $C_E(\tau)$ gives the thermal fluctuation 
of the total energy in grand canonical ensemble, i.e.
\begin{align}
  \frac{\langle \delta \bar{T}_{00}^2 \rangle}V
  = C_E(\tau) ,
  \label{eq:<T2>}
\end{align}
where the value of $\tau$ in this formula must satisfy 
$\tau \gtrsim 2\sqrt{2t}$ in order to avoid the smearing problem.
In the continuum theory the value of $\tau$ in Eq.~(\ref{eq:<T2>}) 
is arbitrary except for $\tau=0$ because of Eq.~(\ref{eq:d_tau}).

Combining Eq.~(\ref{eq:<T2>}) with the linear-response relation 
Eq.~(\ref{eq:c_V}), one can make an estimate of the specific heat 
by reading the value of $C_E(\tau)$ in the plateau region 
in Fig.~\ref{fig:correlator}.
Here, we use the value of $C_E(\tau)$ at midpoint $\tau=\beta/2$.
The specific heat is then obtained as 
\begin{align}
c_V/T^3=15.3(36) \quad {\rm for} \quad T/T_c=2.31.
\label{eq:cvt}
\end{align}
In the analysis of the specific heat of the $SU(3)$ gauge 
theory in Ref.~\cite{Gavai:2004se} using 
the differential method, the values are estimated as 
$c_V/T^3=15(1)$ for $T/T_c=2$ and 
$c_V/T^3=18(2)$ for $T/T_c=3$.
Although we have not taken the continuum extrapolation,
it is notable that our result in Eq.~(\ref{eq:cvt}) is 
consistent with the one in Ref.~\cite{Gavai:2004se}.
This result opens a possibility to measure the specific
heat in a novel method using the linear-response relation
Eq.~(\ref{eq:c_V}).

\section{Summary and outlook}

In this study, we analyzed the expectation values and 
correlation functions of the energy-momentum tensor (EMT)
on the lattice using the EMT operator Eq.~(\ref{eq:T^R}) 
defined by the gradient flow and the small flow time expansion.
Our results show that the numerical analyses of the EMT on the 
lattice with Eq.~(\ref{eq:T^R}) work quite successfully.
In fact, the thermodynamics observables analyzed in our approach 
agree with the previous results using integral method.
Besides, the result on the energy-energy correlator 
is consistent with the conservation law and the linear-response
relation Eq.~(\ref{eq:c_V}).
Moreover, our approach can perform these measurements with 
a significantly small statistics compared 
with the previous methods.

As discussed at the beginning of this manuscript,
the measurement of the EMT on the lattice has been 
a difficult subject, although the EMT is one of the most 
fundamental quantities in physics.
Since the EMT defined by Eq.~(\ref{eq:T^R}) can become 
a solution of this longstanding problem, 
there are a lot of applications of the present study.
Among them, the accurate measurement of the correlation
functions of the EMT is one of the most interesting 
subject.
While we have only presented the numerical result on the 
energy-energy correlator in this manuscript, the same analysis 
can be performed for various channels.
In particular, the correlation functions of the spatial
components of the EMT, $T_{12}$ and $\sum_i T_{ii}$,
are interesting because they are related with shear and bulk 
viscosity, respectively, via Kubo formulas.
The measurements of these correlators with Eq.~(\ref{eq:T^R}) 
thus would enable us a stable analysis of these quantities
on the lattice.
Furthermore, the measurements of the energy and momentum 
distributions in hadrons and flux tube will also be possible 
using the EMT operator in Eq.~(\ref{eq:T^R}).
These analyses will reveal various aspects of the hadron 
structure and confinement.
The application of this method to full QCD with fermion
fields is another important future study 
\cite{Luscher:2013cpa,Makino:2014taa}.
These analyses will be reported elsewhere.

\section*{Acknowledgments}

Numerical simulation for this study was carried out on 
NEC SX-8R and SX-9 at RCNP, Osaka University, and Hitachi 
SR16000 and IBM System Blue Gene Solution at
KEK under its Large-Scale Simulation Program 
(Nos.~T12-04 and 13/14-20). 
The work of M.~A.,
M.~K., and H.~S. are supported in part by a Grant-in-Aid for Scientific
Researches~23540307 and 26400272, 25800148, and 23540330, respectively. 
E.~I. is supported
in part by Strategic Programs for Innovative Research (SPIRE) Field~5. 
T.~H. is partially supported by RIKEN iTHES Project.

\bibliographystyle{JHEP}
\bibliography{ref}

\end{document}